\documentclass[structabstract]{aa}  
\usepackage{graphics,latexsym,amssymb,times,psfig}
\usepackage{txfonts}
\newcommand{\swift}{\emph{Swift~}}

\begin{document}
   \title{Afterglow rebrightenings as a signature of a long-lasting central engine activity? \\The emblematic case of GRB 100814A.
  }
\authorrunning{Nardini et al. 2013}
\titlerunning{The emblematic case of GRB 100814A.}
   \subtitle{ }

   \author{M. Nardini\inst{1}, J. Elliott\inst{2}, R. Filgas\inst{3}, P. Schady\inst{2},  J. Greiner\inst{2}, T. Kr\"uhler\inst{4,5},  S. Klose\inst{6}, P. Afonso\inst{7}, D. A. Kann\inst{2}, A. Nicuesa Guelbenzu\inst{6}, F. Olivares E.\inst{8}, A. Rau\inst{2},  A. Rossi\inst{6}, V. Sudilovsky\inst{2} \and S. Schmidl\inst{6}}

          %\inst{2}\fnmsep\thanks{Just to show the usage
        %  of the elements in the author field}
          %}

   \institute{Universit\`a degli studi di Milano-Bicocca, Piazza della Scienza 3, 20126, Milano, Italy\\
              \email{marco.nardini@unimib.it}
              \and
             Max-Planck-Institut f\"ur extraterrestrische Physik, Giessenbachstrasse 1, 85748 Garching, Germany  
               \and
              Institute of Experimental and Applied Physics, Czech Technical University in Prague, Horska 3a/22, 12800 Prague 2, Czech Republic
              \and
             Dark Cosmology Centre, Niels Bohr Institute, University of Copenhagen, Juliane Maries Vej 30, 2100 Copenhagen, Denmark
              \and
              European Southern Observatory, Alonso de C—rdova 3107, Vitacura, Casilla 19001, Santiago 19, Chile.
              \and
             Th\"uringer Landessternwarte Tautenburg, Sternwarte 5, 07778 Tautenburg, Germany
               \and
               American River College, Physics and Astronomy Dpt., 4700 College Oak Drive, Sacramento, CA 95841, USA
            \and
              Departamento de Ciencias Fisicas, Universidad Andres Bello, Avda. Republica 252, Santiago, Chile               
                \\
  }             
%             \email{}
%             \thanks{The university of heaven temporarily does not
%                     accept e-mails}
           
%\object{GRB 100814A}
   \date{}

% \abstract{}{}{}{}{} 
% 5 {} token are mandatory
 
  \abstract   % context heading (optional)
   {In the past few years the number of well-sampled optical to NIR light curves of long Gamma-Ray Bursts (GRBs) has greatly increased particularly due to simultaneous multi-band imagers such as GROND. Combining these densely sampled ground-based data sets with the \swift UVOT and XRT space observations unveils a much more complex afterglow evolution than what was predicted by the most commonly invoked theoretical models.  GRB 100814A represents a remarkable example of these interesting well-sampled events, showing a prominent late-time rebrightening in the optical to NIR bands and a complex spectral evolution. This represents a unique laboratory to test the different afterglow emission models.
}
   {Here we study the nature of the complex afterglow emission of GRB 100814A in the framework of different theoretical models. Moreover, we compare the late-time chromatic rebrightening with those observed in other  well-sampled long GRBs. }
   {We analysed the optical and NIR observations obtained with the seven-channel Gamma-Ray burst Optical and Near-infrared Detector  at the 2.2 m MPG/ESO telescope together with  the X-ray and UV data detected  by the instruments onboard  the \swift observatory. The broad-band afterglow evolution, achieved by constructing multi-instrument light curves and spectral energy distributions, will be discussed in the framework of different theoretical models.}
   {We find that the standard models that describe the broad-band afterglow emission within the external shock scenario fail to describe the complex evolution of GRB 100814A, and therefore more complex scenarios must be invoked. 
 The  analysis of the very well sampled broad-band light curve of GRB 100814A allowed us to deduce that models invoking late-time activity of the central engine in the observed afterglow emission are the preferred ones for all the different observed features.    This late-time activity most likely has the form of a delayed reactivation of the ejecta emission process. However, a more detailed modelling of the radiative mechanisms associated with these scenarios is necessary to arrive at a firm conclusion on the nature of the optical rebrightenings that so often are detected in long GRBs.  }
   {}


   \maketitle
%
%________________________________________________________________

\section{Introduction}
In recent years, the increasing number of long Gamma-Ray Bursts (GRBs), for which a rich multi-band optical-near infrared (NIR) follow-up is available has, allowed us to discover that the afterglow light curve evolution is anything but simple. The simplistic idea that the optical afterglow can be described as the sum of two or three smoothly connected power-laws produced by synchrotron radiation in the framework of the standard external-shock model (see e.g., Laursen \& Stanek 2003;  Panaitescu et al. 2006; Panaitescu 2007a,b) now fails to explain the complex behaviour observed in a large portion of well-sampled events (see e.g. De Pasquale et al. 2009; Ghisellini et al. 2009; Kong et al. 2010; Nardini et al. 2010, 2011; Filgas et al. 2011a,b; Vlasis et al. 2011). An important role in furnishing the rich datasets necessary for detailed emission mechanism testings is played by instruments such as the Gamma-Ray burst Optical Near-infrared Detector (GROND)\footnote{Tables containing the full GROND and UVOT data samples are only available in electronic form at the CDS via anonymous ftp to cdsarc.u-strasbg.fr (130.79.128.5)
or via {\tt http:\/\/cdsweb.u-strasbg.fr\/cgi-bin\/qcat?J/A+A/}}, a seven-band simultaneous optical to NIR imager mounted on the 2.2 m MPG/ESO telescope at La Silla observatory (Greiner et al. 2008). Combining the simultaneous $g^\prime, r^\prime, i^\prime, z^\prime, J, H, K_s$ GROND observations with UVOT and XRT data gives an extensive and densely sampled picture of the afterglow emission from the NIR to the X-ray bands. This synergy recently allowed us to discover a particularly interesting family of long GRBs, whose light curves are characterised by the presence of prominent chromatic rebrightenings in the GROND bands, for example, GRB 081029 (Nardini et al. 2011) and GRB 100621A (Greiner et al. 2013). These extremely fast rebrightenings cannot be explained by any of the commonly proposed emission models. The growth of the number of events well covered by GROND  shows that GRBs such as GRB 081029 and 100621A can only represent the tip of the iceberg of a more highly populated family of events for which a common additional component needs to be invoked to explain the spectral evolution of the late-time afterglow emission. Other events could share the same emission processes even if the steepness of the optical bump is not that extreme\footnote{Note that optical rebrightenings have been observed since the very beginning of the ``afterglow era''. GRB 970508 (Vietri 1998;
Sokolov et al. 1998) showed a similar bump, but the lack of simultaneous multi-band follow-up did not allow  testing the emission scenario from the spectral point of view.}. GRB 100814A is a remarkable example of these  events. Combining GROND and UVOT data, we compiled a dataset with more than 1300 single photometric detections from 160 to $10^7$ s after the gamma-ray trigger. The same time range is fully covered in the X-rays by XRT observations.  This extremely dense and long-lasting multi-band follow-up allows us to unveil the complex evolution of the afterglow light curve that would have been left hidden in the past when the available data coverage was poorer. The most relevant features in GRB 100814A are an intense X-ray flare in the first 200 s and two rebrightenings at $\sim$ 400 s and 40 ks in the GROND and UVOT bands. 
Moreover, the entire light curve is characterised  throughout by an intense spectral evolution that tracks both the GROND and UVOT light curves very well.

In this paper we first (\S \ref{obs}) present the full available dataset,  then describe the complete broad-band afterglow evolution from the temporal (\S \ref{lc}) and spectral (\S \ref{bbsed}) point of view. We finally discuss (\S \ref{lcphases}) the radiative mechanisms at the basis of the emission in the different phases of the observed light curve in the framework of different theoretical models with particular focus on the nature of the late rebrightening (\S \ref{bump}).

\section{Observations and data reduction}
\label{obs}

\subsection{Prompt gamma-ray observations}
 GRB 100814A (\swift trigger 431605) was detected by the \swift Burst Alert Telescope (BAT) at 03:50:11 UT on 2010 October 14 (Beardmore  
et al. 2010) at the position ${\rm RA}({\rm J}2000)= 22.479\degr = 01^{\rm h} 29^{\rm m} 55^{\rm s} , {\rm Dec.(J}2000) =-17.990\degr = -17\degr 59' 25\farcs7$ (Krimm, et al. 2010). The BAT light curve is characterised by three spikes peaking at $\approx$ 5, 70, and 145 s after the trigger. GRB 100814A is a relatively long GRB with a  duration  $T_{90}(15 - 350 {\rm keV} ) = 174.5 \pm 9.4$ s. This GRB also triggered the Gamma-Ray Burst Monitor (GBM) (Meegan et al. 2009) on board the {\it Fermi} observatory (trigger 303450610 / 100814160) and  Konus-Wind (Aptekar et al. 1995). The GBM  duration is  $T_{90}(50 - 300 {\rm keV} ) = 149 \pm 1$ s, and the time-averaged spectrum is best fitted by a Band function with $E_{\rm peak} = 106.4^{+ 13.9}_{-12.6}$ keV, $\alpha = -0.64^{+0.14}_{-0.12}$, and $\beta = -2.02^{+0.09}_{-0.12}$. The  fluence measured in the GBM energy range  (10-1000 keV)  is $1.98 \pm 0.06\times 10^{-05}~{\rm erg/cm}^2$ and the 1.024-sec peak photon flux  $4.5 \pm 0.2 {\rm ph/s/cm}^2$ (von Kienlin, 2010). The time integrated spectrum as measured by Konus Wind in the 20 keV - 2 MeV energy range  agrees with what is observed by GBM, being well fitted by a power law with an exponential cutoff model with $\alpha  = -0.4^{+0.4}_{-0.3}$, and $E_{\rm peak} = 128^{+23}_{-17}$ keV (Golenetskii et al. 2010). Given the redshift $z=1.44$ measured by O'Meara et al. (2010), the GBM fluence translates into an isotropic energy release of $E_{\rm iso}=1.04\times 10^{53}$ erg and the rest-frame peak energy is $E^{\rm rf}_{\rm peak} = 259^{+34}_{-31}$ keV. With these values GRB 100814A agrees perfectly with the updated version of the Amati (Amati et al. 2002) $E^{\rm rf}_{\rm peak} - E_{\rm iso}$ relation (see Nava et al. 2012).

\subsection{UV, optical, and NIR observations}
\subsubsection{GROND observations and data analysis}
\label{grondobs}
The GROND instrument reacted promptly to the \swift gamma-ray detection notice and started observing GRB 100814A 2.5 min after the trigger alert and 7 min after the GRB trigger.  We immediately detected a bright  afterglow candidate (Filgas et al. 2010) in all seven optical to NIR bands (i.e.; $g^\prime, r^\prime, i^\prime, z^\prime,$ $J,H,K_s$) at the coordinate reported by Schaefer et al. (2010). GROND continued observing GRB 100814A during the first night  as long as it was observable from La Silla ($\sim$ 23 ks) and then densely covered the light curve evolution   for the following three weeks for a total of about 200 individual observations for each band. The GROND  optical and NIR image reduction and photometry were
performed using standard IRAF tasks (Tody 1993) similar to the
procedure described  in Kr\"uhler et al. (2008). A general
model for the point--spread function (PSF) of each image was
constructed using bright field stars and  was then fitted to the afterglow. Optical photometric calibration was performed relative to the
magnitudes of five secondary standards in the GRB field. During photometric conditions, on 2010 November 10, a primary SDSS standard
field (Smith et al. 2002) was observed within a few minutes
of observing  the GRB field. The obtained zeropoints were
corrected for atmospheric extinction and used to calibrate stars
in the GRB field. The apparent magnitudes of the afterglow were
measured with respect to the secondary standards. The absolute calibration of the $J H K_{\rm s}$ bands was obtained
with respect to magnitudes of the Two Micron All Sky Survey
(2MASS) stars within the GRB field  (Skrutskie et al. 2006).

\subsubsection{UVOT observations and data analysis}
\swift observations with the UVOT and XRT narrow-field instruments began ~80s after the BAT trigger and an uncatalogued bright optical source was immediately detected in the first UVOT $u$-and $white$-band finding-chart images (Gronwall \& Saxton 2010). Although undetected in the initial 10s $v,b,w1,m2$ and $w2$ exposures, the afterglow was clearly detected in all seven UVOT optical and ultraviolet (UV) lenticular filters in longer exposures after T+4000s until T+$6\times 10^5$s, after which the afterglow faded below the sensitivity of the UVOT. UVOT photometry was carried out on pipeline-processed sky images downloaded from the {\em Swift} data centre\footnotemark[1] following the standard UVOT procedure (Poole et al. 2008). Source photometric measurements were extracted from the UVOT early-time event data and later imaging data files using the tool {\sc uvotmaghist} (v1.1) with a circular source extraction region of $5\arcsec$ radius for the first 4ks of data, after which a $3.5\arcsec$ source region radius was used to maximise the signal-to-noise ratio. To remain compatible with the effective area calibrations, which are based on $5\arcsec$ aperture photometry (Poole et al. 2008), an aperture correction was applied on the latter, $3.5\arcsec$ source region file photometry.

\subsection{X--ray data reduction and spectral analysis}
\label{xspectrum}
We analysed the XRT data of GRB 100814A with the {\it Swift} 
software package  distributed with 
HEASOFT ({\it v6.12}).  The XRT data were reprocessed  with the  \texttt{XRTPIPELINE} 
tool\footnote{Part of the XRT software,
  distributed with the HEASOFT package:  {\tt http://heasarc.gsfc.nasa.gov/heasoft/}}.  Swift started observing GRB 100814A in windowed timing (WT) mode and moved to photon counting (PC) mode 450 s after the trigger when the photon count rate was $\sim 4$ cts/s. Before $\sim$ 6 ks the PC mode count rate exceeded 0.6  cts/s, therefore a standard pile-up  correction procedure was applied (see Moretti et al. 2005,
Romano et al. 2006; Vaughan et al. 2006). The extraction used a circular region of a typical width of 25  pixels, as discussed in Evans et al. (2009). Background
spectra were extracted in source-free regions of the same size far from the
source. For all of the spectra we created ancillary response
files (ARF) with the xrtmkarf tool and used the calibration
database updated to March 2012. The spectra were re-binned
to have a minimum of 20 counts per energy bin, and energy
channels below 0.3 keV and above 10 keV were excluded
from the analysis. The XSPEC (v12.7.1) software was used
for the analysis.
 
In this paper we focus on the afterglow emission of GRB 100814A and, since the WT mode observations are dominated by the last spike of the prompt gamma-ray emission and by the curvature effect steep-decay phase, we limit our spectral analysis to the PC mode data. We extracted three spectra integrated during  different light curve phases. The spectral analysis results are reported in Table \ref{spectrumxrt}.

 \begin{table}
 \caption{Best-fit parameters for the XRT spectral analysis}
\begin{center}
\begin{tabular}{c c c c c c}
\hline
\hline
$t_{\rm start}$ - $t_{\rm end}$&  exposure & $N_{\rm H}^{\rm host}$  & $\beta_{\rm X}$ & Cstat & dof\\
$[ks]$ 	                                & $[ks]$ 	  			&$[10^{21}cm^{-2}]$                   &     & &\\
\hline
\hline
3.9-12.2	&2.2&	$5.3^{+12}_{-5.3}$&	$0.82\pm 0.11$	&  273	&346\\
21.2-145.2	&28.4	&$9.8\pm 5.2$&	$0.91\pm 0.05$&	578	&574\\
212.0-1000	&73.1	&$24\pm 15$		&$1.03\pm 0.13$	&307&	320\\
\hline
\hline
\end{tabular}
\tablefoot{Results of the X--ray spectral fitting.  We report
 the time interval in which  the  spectrum was extracted, the equivalent neutral hydrogen
  column density at the host redshift  $N_{\rm H}^{\rm host}$, the unabsorbed spectral index $\beta_{\rm X}$,   the Cash statistics parameter Cstat (Cash 1979), and the number of degrees of freedom.
  }
  \label{spectrumxrt}
\end{center}
\end{table}

\section{Afterglow temporal evolution}
\label{lc}
\subsection{X-ray light curve}
\label{lcx}
\label{X}
\begin{figure}
\resizebox{\hsize}{!}{\includegraphics{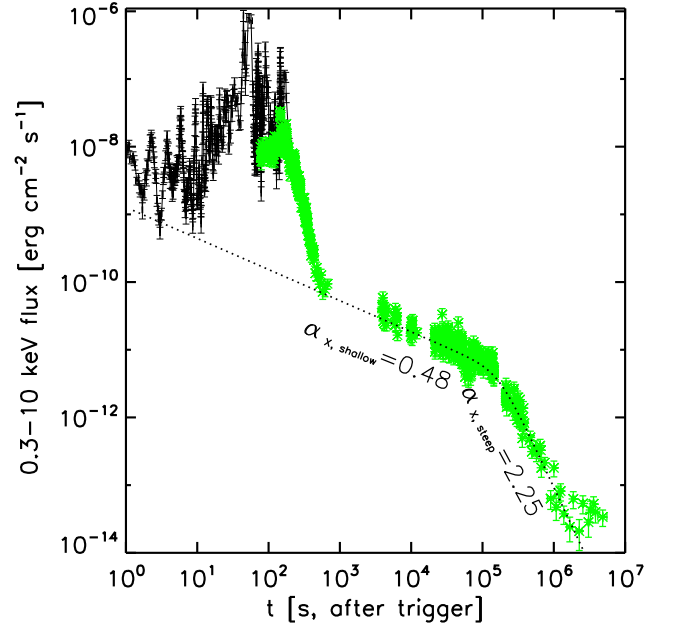}}
\caption{Observed BAT (black line) and XRT (green crosses) light curves of GRB100814A rescaled at 0.3-10 keV superposed on the late-time XRT afterglow  best-fit model described in \S \ref{lcx}. Data taken from the \swift burst analyser {\tt http://www.swift.ac.uk/burst\_analyser/00431605/ (Evans et al. 2010)}.
}
\label{lcgx}
\end{figure} 

The fast spacecraft slew, together with the relatively long GRB duration, allowed XRT to start observing GRB 100814A while the prompt gamma-ray emission was still detected by BAT (87 s after the trigger), as shown in Fig. \ref{lcgx}. Between 90 and 200 s, the XRT light curve brightened by a factor of $\sim$ 5. This flare  was simultaneous to the last FRED pulse observed in the BAT band which peaked at 145s (Saxton et al. 2010). After the peak, the X-ray flux faded by a factor of $\sim 150$ in about 300 s and a very long shallow decay phase began. Between 550 s and  130 ks the light curve can be described by a single power-law decline. Using the standard notation $F(\nu, t)\propto t^{-\alpha}\nu^{-\beta}$, we have during this shallow phase $\alpha_{\rm X, shallow}= 0.52 \pm 0.05$ ($\chi^2_{\rm red}=1.46$). The main reason for the high value of $\chi^2_{\rm red}$ is  a clear flux fluctuation below the best-fit power-law  evolution between 55 and 75 ks. The shallow decay phase ended with a sudden steepening of the light curve occurring at around 130 ks followed by a faster decay phase lasting until 1 Ms. Fitting the XRT light curve in the interval between 500 and $10^6$ s with a smoothly connected broken power-law with the parametrisation from Beuermann et al. (1999), we obtain $\alpha_{\rm X, shallow}= 0.48 \pm 0.02$,  $\alpha_{\rm X, steep}= 2.1 \pm 0.1$ and a break time at $173\pm 5.8$ ks ($\chi^2_{\rm red}=1.39$). The highest residuals are again represented by the flux decrease around 60 ks. 

After $2\times 10^6$ s the XRT flux remained constant. We obtained a dedicated Chandra observation (Obs-ID 15493)
 to investigate this unusually persistent X-ray flux. The observation
 was obtained on  2012 October 22, and lasted 38.5 ks (01:11-12:46 UT).
 We clearly detect an X-ray source at ($4.7\pm0.1) \times 10^{-3}$ cts/s,
 consistent with the Swift/XRT rate of $0.7 \times 10^{-3}$ cts/s (taking 
 into account the different sensitivities). However, the position of
 the X-ray source is RA(2000.0)=02:39:53.96, Decl.(2000.0)=-27:59:35.5
 with an error of $\pm$0.5 arcsec. This is 9.8 arcsec offset from the
 GRB host position and thus incompatible with an association to 
 GRB 100814A. The X-ray position coincides with an optical source
 seen in the GROND images, and thus is clearly an unrelated source
 with its own optical counterpart. We therefore interpret the
 constant X-ray flux in the Swift/XRT light curve as generated by this
 nearby source, and unrelated to GRB 100814A, and ignore it from here on.   A fit to the XRT data that also accounts
for the contamination from this unrelated source gives a steeper value
for  $\alpha_{\rm X, steep}= 2.25 \pm 0.1$, as shown in Fig. \ref{lcgx}. 

\subsection{GROND light curve}
\label{GRONDlightcurve}
\begin{figure}
\resizebox{\hsize}{!}{\includegraphics{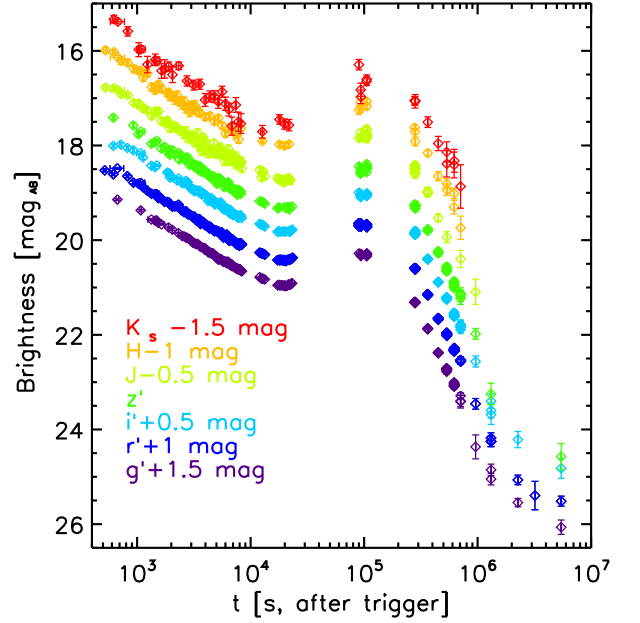}}
\caption{Observed GROND seven-band light curve  of the afterglow of GRB 100814A. A constant shift has been applied to the different bands for clarity, as reported in the label. Fluxes were not corrected for Galactic foreground extinction. The full GROND data set is available as on-line material.
}
\label{lcopt}
\end{figure}

The GROND light curve is characterised by at least three different phases, as shown in Fig \ref{lcopt}.  From 600 s to about 10 ks the evolution can be  described in all  seven bands by a single shallow  power-law decay with an index  $\alpha_{1, \rm opt}=0.57\pm 0.02$ ($\chi^2_{\rm red}=1.22$ obtained by fitting all seven bands simultaneously). A weak fluctuation is present in the first three data, but no additional component is needed in the GROND-only sample.  The value of $\alpha_{1, \rm opt}$ is comparable to the one observed in the X-rays in the same time interval. The decay rate started to decrease 10 ks after the trigger, showing a weak rebrightening at the end of the first night around $\sim 20$ ks after the trigger. This flux rise was confirmed by the second night's observations between 90 and 105 ks. Unlike the first night evolution, the rebrightening was strongly chromatic with the $g^\prime$-band flux increasing by a factor of $\sim 3.2$, while $K_{s}$ brightened by a factor of $\sim 4$.  As a first approximation, we fitted each band's light curve between 20 and 100 ks, with a single power-law $F\propto t^{-\alpha_{band, reb}}$ ,obtaining
\begin{itemize}
\item $\alpha_{\rm g^\prime, reb}=-0.40\pm 0.02$ ($\chi^2_{\rm red}=1.8$ with 11 d.o.f.)
\item $\alpha_{\rm r^\prime, reb}=-0.46\pm 0.01$ ($\chi^2_{\rm red}=1.3$ with 23 d.o.f)
\item  $\alpha_{\rm i^\prime, reb}=-0.49\pm 0.02$ ($\chi^2_{\rm red}=1.2$ with 11 d.o.f.)
\item  $\alpha_{\rm z^\prime, reb}=-0.50\pm 0.02$ ($\chi^2_{\rm red}=1.6$ with 12 d.o.f.)
\item $\alpha_{\rm J, reb}=-0.58\pm 0.03$ ($\chi^2_{\rm red}=1.6$ with 6 d.o.f.)
\item $\alpha_{\rm H, reb}=-0.53\pm 0.06$ ($\chi^2_{\rm red}=1.0$ with 3 d.o.f.)
\item $\alpha_{\rm K_s, reb}=-0.55\pm 0.09$ ($\chi^2_{\rm red}=0.7$ with 3 d.o.f.).
\end{itemize}
This confirms the chromatic nature of the bump. The significantly large $\chi^2_{\rm red}$ obtained in the optical is most likely due to an intrinsic fluctuation during the second-night observations that is statistically larger than the photometric uncertainty; this feature is discussed in \S \ref{bump}.
After this rebrightening, the optical to NIR afterglow of GRB 100814A entered a third phase characterised by a steep decay; this phase lasted until $10^6$ s when the contribution of the host galaxy became dominant. This phase was also characterised by a chromatic evolution during which  the bluer  bands decayed  faster than the NIR bands.   Separate fits for the different bands from $2\times10^5$ to $10^6$ s give
\begin{itemize}
\item $\alpha_{\rm steep, g^\prime}=2.03\pm 0.02$ ($\chi^2_{\rm red}=1.5$ with 43 d.o.f)
\item $\alpha_{\rm steep, r^\prime}=1.95\pm 0.02$ ($\chi^2_{\rm red}=1.8$ with 48 d.o.f)
\item  $\alpha_{\rm steep, i^\prime}=1.97\pm 0.02$ ($\chi^2_{\rm red}=0.53$ with 43 d.o.f)
\item $\alpha_{\rm steep, z^\prime}=1.92\pm 0.03$ ($\chi^2_{\rm red}=0.53$ with 41 d.o.f.)
\item  $\alpha_{\rm steep, J}=1.91\pm 0.08$ ($\chi^2_{\rm red}=0.58$ with 18 d.o.f.)
\item $\alpha_{\rm steep, H}=1.80\pm 0.09$ ($\chi^2_{\rm red}=0.97$ with 8 d.o.f.)
\item  $\alpha_{\rm steep, K_{\rm s}}=1.65\pm 0.14$ ($\chi^2_{\rm red}=0.52$ with 8 d.o.f.).
\end{itemize}

After $10^6$ s the GROND light curve was dominated by the contribution of the underlying host galaxy that is clearly detected in the $g^\prime, r^\prime, i^\prime, z^\prime$ optical bands. Correcting the optical magnitudes for the contribution of the host galaxy, the chromaticity of the steep decay loses significance, and the decay of all optical bands is consistent with a unique value $\alpha_{\rm steep, opt}=2.25\pm 0.08$. This probably occurs also in the the $J, H$, and $K_{s}$ bands but  in these bands the host galaxy was too faint to be detected and therefore no conclusive result can be obtained for  the NIR bands.

\subsection{UVOT light curve}
\label{uvot}
\begin{figure}
\resizebox{\hsize}{!}{\includegraphics{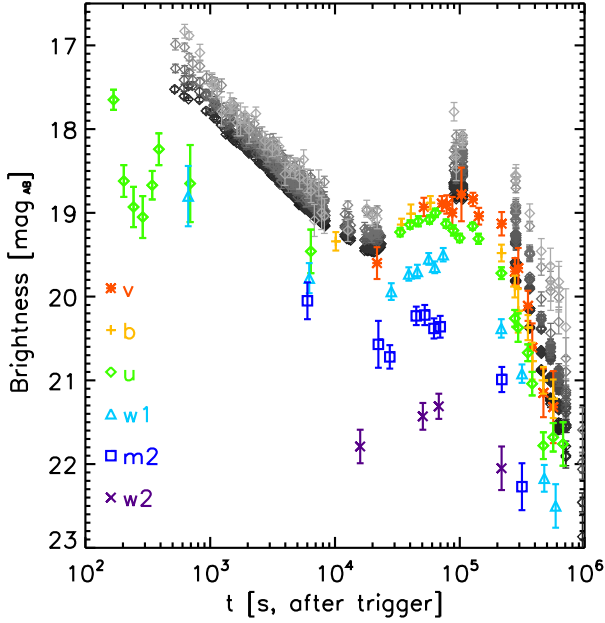}}
\caption{UVOT light curve (coloured dots) of the afterglow of GRB 100814A superposed on the GROND dataset (grey dots) plotted in Fig. \ref{lcopt}. 
Magnitudes  were not corrected for Galactic foreground extinction. The full UVOT data set is available as on-line material.
}
\label{luvot}
\end{figure}

Like for the XRT, UVOT observations started while the prompt gamma-ray phase was still detected by BAT. The first 3 $u$ band magnitudes plotted in Fig. \ref{luvot}  are therefore simultaneous to the decay phase of the bright X-ray flare discussed in \S \ref{X}.  Between 300 and 600 s after the trigger, the $u$ band light curve showed a prominent rebrightening of about 0.8 mag that  hints at the peak seen in the first three GROND  observations (see \S \ref{GRONDlightcurve}). Considering the GROND and UVOT data sets together, we can therefore confirm the light curve peak at $630\pm40$s. During the smooth decay-phase observed by GROND between 640 and 10 ks, UVOT observed in $b$, $u$, $w1$ and $m2$. Alhough much more sparse and affected by larger uncertainties, the UVOT temporal evolution is consistent with the single power-law decay with   $\alpha_{1, \rm opt}=0.57\pm 0.02$ measured with GROND. The UVOT coverage becomes fundamental between 25 and 80 ks after the trigger (corresponding to the daytime between the first and second night in Chile) when no GROND observation is available. The rising phase of the strong late-time optical rebrightening is clearly tracked in all UVOT bands. As can be seen in Fig. \ref{luvot}, the fact that this rebrightening is more prominent at redder wavelengths is confirmed. Moreover, thanks to the rich coverage and the small uncertainties in the $v$, and $u$ bands at $\sim$ 60 ks, we can identify the  smaller-scale fluctuations around the rebrightening peak. Those wiggles are similar to the ones observed around the peak of the prominent optical rebrightenings of  GRB 081029 and GRB 100621A  discussed in Nardini et al. (2011) and Greiner et al. (2013), respectively. 
\\
The late fading in the UVOT bands tracks  the power-law decay observed in the bluer GROND bands well. The larger uncertainties that affect the UVOT data compared with GROND do not allow us to see the differences among the bands. Fitting together $v, b, u, w1$, and $m2$ light curves from $2\times 10^5$ to $10^6$ s  with a single power-law decay, we obtain $\alpha_{\rm steep, UVOT}=2.08\pm 0.09$ ($\chi^2_{\rm red}=0.92$), consistent with the decay seen in the optical GROND bands, but not with that seen in the NIR bands. No correction for the host galaxy contribution is possible.

\label{grondlc}

\subsection{Complete UV-NIR light-curve}
\label{bblc}

Since the light curve  became strongly chromatic at the rebrightening and onwards, it is not possible to describe the full multi-colour afterglow evolution with a unique functional form. We decided to fit the monochromatic evolution as a combination of two separate but coexisting components, as was done, for example,  for GRB 081029  (Nardini et al. 2011). Compared with GRB 081029,  GRB 100814A requires a more complex functional form, that is the sum of two smoothly connected triple power-laws plus the contribution of the host Galaxy becoming dominant in the optical bands  after $\sim 10^6$s. 

\begin{figure}
\resizebox{\hsize}{!}{\includegraphics{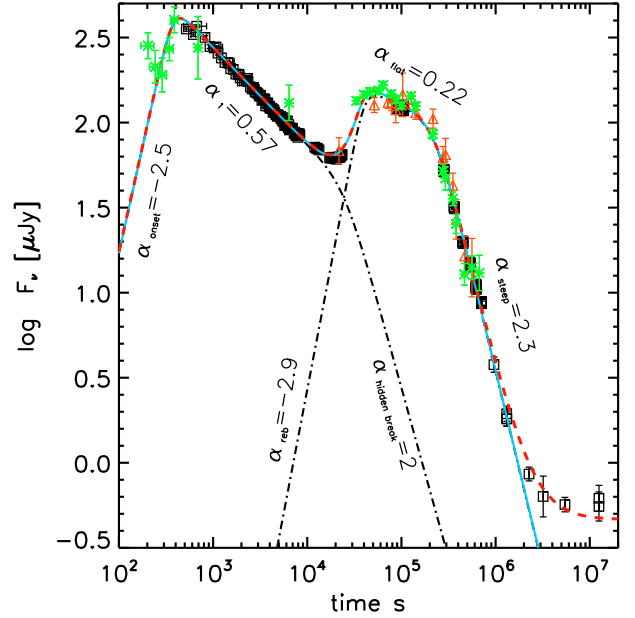}}
\caption{Monochromatic $r^{\prime}$ band light curve of GRB 100814A fitted as the sum of two smoothly connected triple power-laws plus the contribution of the host galaxy (red dashed line). Black squares represent the GROND  $r^{\prime}$ photometry, while green crosses and red triangles represent  UVOT $u$ and
$v$ bands respectively, rescaled to the  $r^{\prime}$ band flux as described in the text.  }
\label{fittotale}
\end{figure}

In this section we describe the fitting of the GROND $r^{\prime}$ band because it has better-quality  photometry (i.e high dataset density and small photometric errors).  
 All other bands can be well fitted with the same model.  However, we note
that the best-fit values of the rebrightening and steep decay slopes determined with this two-component model differ from  those determined in Sect. 3.2., where we
fitted each component alone (i.e., the result of the degeneracy
between  $\alpha_{1}$, $\alpha_{\rm reb}$, and $\alpha_{\rm hb}$ where $_{\rm hb}$ means hidden break). In particular, this makes $\alpha_{\rm reb}$ much steeper than what was found only fitting the bump.
 Since the initial afterglow onset and the rebrightening phase were only covered by the UVOT observations, we exploit the information brought by these bluer bands to better constrain the $r^{\prime}$-band fitting at early time. The multi-colour evolution before the rebrightening was achromatic, therefore we normalised the UVOT $u$ and $v$  bands to the simultaneous   $r^{\prime}$-band flux  in that phase as shown in Fig. \ref{fittotale}. A different normalisation was used after 20 ks  because of  the strong chromaticity of the rebrightening. 

As reported in Fig. \ref{fittotale}, the first two power-laws of the first component  (i.e., $\alpha_{\rm onset}$ and $\alpha_1$)  perfect agree with the results reported in previous sections. After the rebrightening - at $t >500$ ks - the flux decayed  below the
extrapolation of the earlier shallow phase, with $\alpha_1=0.57$. This
means that the first component of emission must, at some point
after the rebrightening, enter a steeper decay phase. We refer to this
assumed steepening as the hidden break. Of course, without any
direct evidence of the exact location of this break, there exists an
obvious degeneracy between the time of the hidden break, $t_{\rm hb}$, its
slope $\alpha_{\rm hb}$, and the slope of the rising part of the second
component, $\alpha_{\rm reb}$. However, we have an observational constraint that puts the break after the starting of the rebrightening, since the previous
light curve is well described by a single power-law. Moving the break to
later times would imply steeper values for both the second-component rise
and the declining phase of the first component. Setting the hidden break
at about 25ks represents the most conservative choice.  Freezing  the break time  at $25\pm 3$ ks, we obtain a post-break decline $\alpha_{\rm hb}\approx 2$ and $\alpha_{\rm reb }= 2.9$.  
The second component is affected by larger uncertainties. Using UVOT observations to constrain the rebrightening peak time that was not covered by GROND, we can estimate a peak of the bump at  42 ks. Considering a persisting contribution of the first component and combining GROND and UVOT coverage, we obtain a  slope for the rising phase  that is steeper than that obtained in \S \ref{GRONDlightcurve}. The value reported in   Fig. \ref{fittotale} is correlated with the value of $\alpha_{\rm hb}$ we assumed; in all cases a value of $\alpha_{\rm reb} >2$ is expected unless an immediate shut-off of the first component is assumed. 

The light curve after $\approx$ 20 ks was  dominated by the second smoothly connected triple power-law with the indices reported in Fig. \ref{fittotale} plus the contribution of the host galaxy. The two breaks are located at 40 and 225 ks, respectively.  The fit of the shallow phase after the peak is  affected by the higher residuals because of the previously mentioned statistically significant fluctuations that are tracked by the UVOT data. The value of $\alpha_{\rm steep}=2.30\pm 0.05$ is consistent with that found for the optical bands in \S \ref{GRONDlightcurve} when considering the contribution of the host galaxy. 

We note that calculating the rest-frame optical luminosity at 12 h rest frame as described in Nardini et al. (2006, 2008a,b) from the observed GROND light curve places GRB 100814A among the intrinsically more luminous afterglows ($\log{L(\nu_{\rm R})^{\rm 12h~rf}}\sim 30.9$) while  extrapolating the first component without considering the contribution of the rebrightening component places GRB 100814A in the sub-luminous family ($\log{L(\nu_{\rm R})^{\rm 12h~rf}}\sim 29.2$) similarly to what observed in the case of GRB 970508 (Nardini et al. 2006).

\section{Broad-band afterglow spectral evolution}
\label{sed}
\subsection{Optical to NIR spectral energy distribution  evolution}
\label{sedopt}
Since GROND observations are simultaneous in all seven bands, we extracted one spectral energy distribution (SED) for each exposure. Data were corrected for the Galactic foreground extinction of E(B-V)=0.02 in the direction of the burst (Schlegel et al. 1998) and then fitted assuming a simple power-law spectrum. The  Large and Small Magellanic Clouds (LMC, SMC) and Milky Way (MW) extinction laws from Pei (1992) were used to describe the dust reddening in the host galaxy. We found that all individual SEDs are consistent with a negligible host-galaxy dust reddening ($A_{\rm V}^{\rm host}<0.07$) regardless of the extinction curve. We therefore fixed  $A_{\rm V}^{\rm host}=0$ and calculated the spectral index for all individual GROND exposures. 

\begin{figure}
\resizebox{\hsize}{!}{\includegraphics{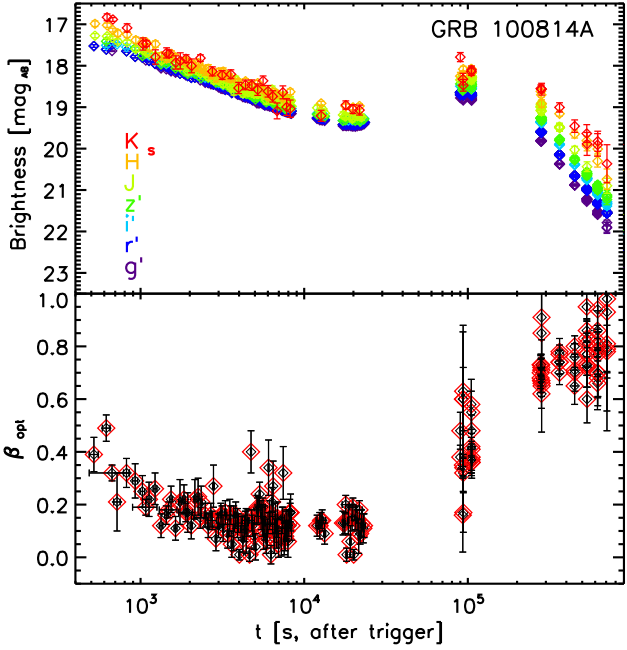}}
\caption{Top panel: GROND light curve reported in Fig. \ref{lcopt}. Bottom panel: Spectral index $\beta_{\rm opt}$ obtained  by fitting each GROND observation with a simple power-law model  after correcting the magnitudes for the  expected Galactic foreground extinction and assuming a host galaxy $A_{\rm V}^{host}=0$ (see \S \ref{sed} for more details). }
\label{beta}
\end{figure}

In Fig. \ref{beta} we show the evolution of $\beta_{\rm opt}$ with time together with the GROND multi-colour light curve. We immediately notice that   $\beta_{\rm opt}$ evolves in time, and closely tracks the flux evolution. During the early smooth decay, the spectral index remains  constant apart from a fluctuation around the first two points that correspond to the light curve peak discussed in \S\ref{uvot}. During this phase the SED, which has a $\beta_{\rm opt}=0.18\pm0.08$, is  bluer than the values tipically obtained in GRB afterglows (see, e.g., Kann et al. 2010).  During the rebrightening at $\sim 10^5$s,  the spectrum evolves and becomes redder,  as anticipated in \S\ref{grondlc}, with  $\beta_{\rm opt}\approx0.5$. The spectral evolution continued to follow this trend between 100 and 500 ks, and the spectral index becomes $\beta_{\rm opt}=0.82\pm0.15$ around $7\times 10^5$s. The larger uncertainties at later times are due to the required correction for the host galaxy contribution. 

Adding UVOT data to the optical SED fits does not affect the GROND-only results. This is  because of the larger intrinsic error on UVOT photometry combined with the systematic uncertainty that must be  added to account for  the non-simultaneity of the observations. Both the negligible $A_{\rm V}^{\rm host}$ and $\beta_{\rm opt}$ evolution are therefore confirmed.

\subsection{Near-infrared to X-ray SED}
\label{bbsed}
\begin{figure}
\resizebox{\hsize}{!}{\includegraphics{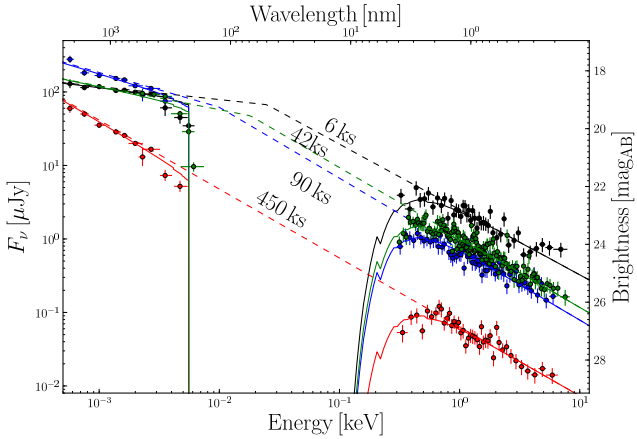}}
\caption{GROND to XRT SED of GRB 100814A extracted in four different time slices as reported in the labels. The details of the fitting procedure are described in \S \ref{bbsed} and the best-fit parameters are reported in Table \ref{bbresult}. Note that bluer UVOT bands are affected by the Ly-alpha absorption at z=1.44. 
}
\label{sedplot}
\end{figure}

We selected three time slices in which quasi-simultaneous GROND, UVOT, and XRT observations are available.  These SEDs are centred at 6, 90, and 450 ks, respectively. We moreover extracted an additional SED at 42 ks, corresponding to the rebrightening phase, which is covered by all six UVOT filters.  According  to the light curve fitting described in \S \ref{bblc}, at the selected times one component (i.e. early- or late-time triple power-law)  strongly dominates the light curve and we therefore did not need to separate the different components contribution.  
\\
Following the procedure described in Greiner et al. (2011),  we extracted the corresponding broad-band SEDs as shown in Fig. \ref{sedplot}. We applied Eq. 5 from Kr\"uhler et al. (2011) to evaluate the amplitude of a possible systematic offset between UVOT and GROND data. Since the result is consistent within 2 sigma for all SEDs, we did not apply any correction to the different data sets and conclude that the instrumental cross-calibration is reliable within measurement uncertainties. The first three SEDs, extracted at 6, 42  and 90 ks respectively,  are inconsistent with all bands laying on the same power-law spectrum (reduced $\chi^2$ larger than 2). We therefore used a broken-power-law model.  We simultaneously fitted the four SEDs with a constant value for $A_{\rm V}^{\rm host}$ and $n_{\rm H}^{\rm host}$ column density.
Assuming SMC-like host galaxy extinction curve, we obtain $A_{\rm V}^{\rm host}<0.04$, which agrees with the GROND-only SED  results and  $n_{\rm H}^{\rm host}=1.3\pm 0.3\times 10^{21}$ cm$^{-2}$, in agreement with the values reported in Table \ref{spectrumxrt}.
\\
The fourth SED, extracted at 450 ks after the trigger,  is instead consistent with a single power-law connecting all bands. This agrees with the fact that at late times, the GROND-only value of $\beta_{\rm opt}$ is consistent with the XRT-only best-fit value of $\beta_{\rm X}$.
The best-fit parameters for the four time slices are reported in Table \ref{bbresult}. We obtained an good fit ($\chi^2_{\rm red}=1.17$ with 301 dof). Under the assumption that the UVOT and X-ray bands have been produced by the same radiative mechanism, the break energy is first located between the optical and XRT bands,  then moves redwards with time, surpassing the optical bands between the third and fourth SED extraction time (i.e., between 90 and 450 ks). As noted in \S \ref{sedopt}, the early optical spectrum is quite blue and the difference between the optical and X-ray spectral indices $\Delta(\beta)=0.71$ is large with respect to the standard values measured in GRB afterglows (Kann et al. 2010). In particular, it is larger (2$\sigma$) than $\Delta(\beta)=0.5$ which we would expect if the break frequency were interpreted as the cooling frequency in a synchrotron-dominated emission scenario. The observed difference decreases in the second SED, becoming exactly 0.5 at 90 ks,  and then becoming consistent with zero after the peak of the light curve. The prominent spectral evolution previously found for the XRT- and GROND-only analysis is confirmed. 
\begin{table}
\caption{Best-fit parameters for the broad-band NIR-to-X-ray SEDs}
\begin{center}
\begin{tabular}{c c c c}
\hline
\hline
Time   & $\beta_{\rm opt}$ & $\beta_{\rm X}$ & Break energy\\
$[ks]$ &                                &                            & $[keV]$\\
\hline
\hline
6.0  & $0.18\pm 0.07$& $0.89\pm 0.05$& $0.025\pm 0.009$\\
42   & $0.32\pm0.19$ &$ 0.95\pm0.03$& $0.018\pm0.010$\\
90   & $ 0.47\pm 0.05$ & $0.97\pm0.05$& $0.01\pm 0.007$\\
450 & $0.94\pm 0.07$& $1.08\pm 0.11$ &---\\
\hline
\hline
\end{tabular}
\tablefoot{Results obtained from the broad-band spectral fitting of GROND+UVOT+XRT observations described in \S \ref{bbsed}. Columns report: the time at which the SEDs were extracted; the best-fit value of  the low-energy spectral index, the  high-energy spectral index,  and the break energy in units of keV.  
  }
  \label{bbresult}
\end{center}
\end{table} 

\begin{figure}
\resizebox{\hsize}{!}{\includegraphics{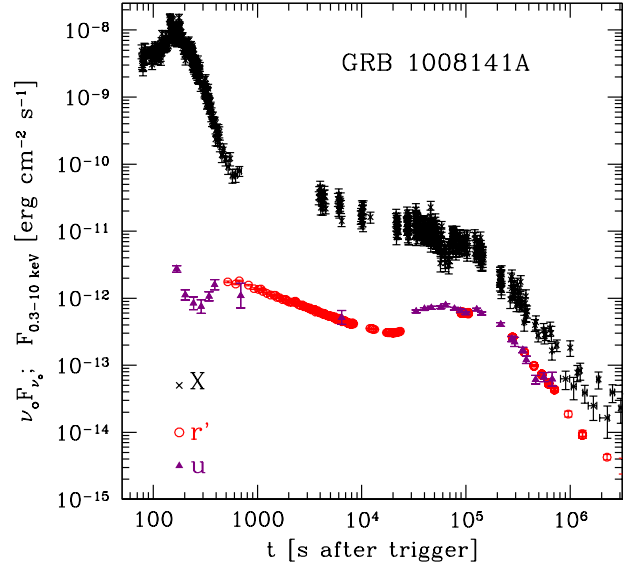}}
\caption{GROND $r^{\prime}$ band (red empty circles) and  UVOT $u$ band (violet filled triangles) $\nu F_{\nu}$ light curve superposed on the unabsorbed 0.3--10 keV XRT light curve (black crosses) of the afterglow of GRB 100814A. UVOT and GROND data were corrected for the Galactic foreground  extinction only.
}
\label{lcobs}
\end{figure}

\section{Different light curve phases}
\label{lcphases}
The complex broad-band evolution of the afterglow of GRB 100814A reveals several components that are dominant at different times. 
\\
For simplicity, we can divide the light curve into five phases:
\begin{itemize}
\item[i)] before 250 s: prompt-afterglow phase;
\item[ii)] $250<t<640$ s: early optical rise;
\item[iii)]  $640<t<2\times 10^4$ s: shallow-decay phase;
\item[iv)]  $2\times 10^4<t<10^5$ s: late rebrightening;
\item[v)]  $10^5<t<2\times 10^6$ s: late steep decline.
\end{itemize}
\subsection{Prompt afterglow}
During phase i) the available optical information is carried by  the first UVOT $u$ band data only, therefore  we cannot constrain the spectral shape. This early bright phase is however simultaneous to the bright X-ray flare discussed in \S\ref{X}. The  last prompt gamma-ray spike detected by BAT 150 s after the trigger, together with the very hard XRT spectrum (photon index from the XRT $\Gamma_{X}\approx 1$ at 150s)\footnote{See the on-line \swift Burst Analyser page for GRB 100814A at {\tt http://www.swift.ac.uk/burst\_analyser/00431605/}}  in this phase, suggests that all \swift data from UVOT to BAT before 250 s are still dominated by the prompt emission. This idea was previously suggested for other promptly detected optical-UV GRBs (see, e.g., the case of the naked-eye burst Racusin et al. 2008; Beskin et al. 2010), but in this case the data sample is not dense enough to constrain a possible delay between gamma-ray and UVOT peak. 

\subsection{Early afterglow: hint of an onset} 
\label{onset}
During phase ii)  optical and X-ray emissions are dominated by different processes. In the X-rays we observe the typical steep decline
 ($\alpha_{\rm X, 1}\approx 5.4$, Saxton et al. 2010), ending $\sim$520 s after the trigger. This steep decline is usually ascribed to the curvature emission of the fireball at the end of the prompt phase (Nousek et al. 2006; Zhang et al. 2006). In the meantime, the optical and UVOT light curves show a rise peaking at about 600 s. The steepness of the rise $\alpha_{\rm onset}=-2.5\pm1.2$ can only be constrained by a few data points, but agrees with the typical values obtained for the pre-peak rise in the so-called fast-rising optical afterglows (FROAs, Panaitescu \& Vestrand 2008 $[\rm{PV08}]$). Considering a peak time at 640s after trigger, we see that GRB 100814A  agrees perfectly with the $F_{\rm peak}^{\rm z=2}-t_{\rm peak}^{\rm z=2}$ correlation between the  z = 2 optical light curve peak flux (at 2 eV) and the peak time found by PV08. These authors claimed that such a correlation is in agreement with a scenario where the observed emission is due to  the synchrotron emission from a collimated blast-wave and for observer locations just outside the jet aperture. In this scenario the rise of the afterglow light curve is caused by the widening of the cone of the relativistically beamed jet emission, as the jet decelerates progressively. With  $F_{\rm peak}^{\rm z=2}=0.28$ Jy and $t_{\rm peak}^{\rm z=2}=787$s, GRB 100814A lies exactly on the best-fit correlation in their Fig. 2.  If the ejecta shell is uniform, the fast rise of the light curve from the pre-decelaration forward-shock implies that the shell is expanding in an homogeneous circumburst medium and the  reverse shock is semi-relativistic (PV08). 
   
\subsection{Post-peak evolution: an achromatic shallow decay}
\label{shallow}
Phase iii) is characterised by an achromatic evolution in the GROND and in the UVOT bands. Moreover,  optical-UV and X-ray bands are well described by a similar power-law  decay with $\alpha_{\rm X, shallow}$ consistent with $\alpha_{1, \rm opt}$ within 2 $\sigma$. This  evidence suggests a common origin for the optical and X-ray emission observed during phase iii). The  quite a shallow ($\alpha_{1, \rm opt}\approx 0.6$) decay  phase after the onset peak is not expected when, as argued in \S \ref{onset}, the ejecta are expanding in a homogeneous medium. PV08 found that such an evolution can be observed in  the case of a structured outflow seen off-axis (right panel of Fig. 3 in PV08, Panaitescu, M\'esz\'aros \& Rees 1998; Granot et al. 2002). Alternatively, a continuous energy injection into the fireball can be invoked.
In \S \ref{bbsed} we have shown that optical and X-ray bands do not belong to the same spectral component (i.e., do not lie in the same power-law fit). The $\Delta (\beta)=0.71$ measured for the first broad-band SED at 6 ks  is only marginally consistent (at $\sim$ 2$\sigma$ level) with the commonly made assumption that the break between the UVOT and X-ray bands is the synchrotron cooling frequency $\nu_{\rm c }$. GRB 100814A  shows one of the bluest optical afterglows ever observed in a long GRB (see Kann et.al., 2010).  In the standard forward-shock model, if we accept that the break
frequency between UVOT and XRT is $\nu_{\rm c }$ (an already marginally
consistent scenario), then the XRT light-curve should decay faster
than the UVOT light-curve, for the ISM environment suggested by the
steep onset of the UVOT light-curve (c.f. \S \ref{onset}). The fact that in this case the optical and X-ray light curves evolve achromatically  therefore disagrees with this scenario. However, in this scenario, the cooling frequency depends on the fraction of the total energy in the magnetic field as $\varepsilon_{\rm B}^{-\frac{3}{2}}$. Usually, a different behaviour of the observed $\nu_{c}$ evolution has been justified by invoking an evolution of the microphysical parameter
$\epsilon_{B}$ (see e.g. Panaitescu et al. 2006; Kong et al. 2010, Filgas
et al. 2011), and this might hold also in this case.

\subsection{Late rebrightening}
\label{bump}
The achromatic evolution that characterises the early afterglow phase ends while the GROND and UVOT light curves started to brighten 20 ks after the trigger. Unlike for GRB 081029 (Nardini et al. 2011) or 100621A (Greiner et al. 2013), GROND was unable to follow the full rising phase since it occurred during daytime in Chile. This temporal window was in part covered by UVOT. As can be appreciated in Figs. \ref{luvot} and \ref{lcobs}, UVOT data in the GROND light curve gap show that  the rising phase had reached  a first peak at 55 ks. This peak was followed by a flat evolution with at least two smaller-scale fluctuations superposed (clearly visible in particular in $u$ and $v$ bands), ending with the start of the steep decline defined as phase iv). These deviations from a simple power-law evolution of the shallow post-rise phase could be either treated as fluctuations as in the case of GRB 081029 and 100621A or could be the signature of the existence of two separate peaks, one at about 40 ks and the other at about 150 ks, whose superposition results in the observed overall rebrightening. This second scenario, invoking two different components in the rebrightening phase, moreover agrees with the clear optical colour change shown in Fig. \ref{beta} between 40 and 200 ks, but the lack of a dense GROND photometric coverage of the whole phase does not allow us to study this scenario in more detail. 
 
Combining the GROND and UVOT photometry, assuming the contribution of two separate components as described in section 3.4, we find the best-fit optical rise  between 20ks and 100ks after the GRB trigger to be $\alpha_{\rm reb}\sim -2.9$. This high value is less extreme than $\alpha_{\rm reb}\sim -8.2$ and $\alpha_{\rm reb}\sim -14$ measured for GRB 081029 (Nardini et al. 2011) and GRB 100621A (Nardini et al. 2012, Greiner et al. 2013), respectively, and agrees with both the typical values for FROAs   reported (PV08) and with the late-bump rise steepnesses derived by Liang et al. (2013) in their sample. Liang et al. (2013) also found a good correlation between the optical luminosity at the rebrightening peak and the host  frame  peak time. GRB 100814A has a $L_{\rm R, P}=6.8\times 10^{45}$ erg s$^{-1}$ at a host-frame peak time of about 17 ks these values  perfectly agree with this correlation (see Fig. 7 in Liang et al. 2013).

 The rebrightening observed from the NIR to the UVOT bands is not visible in the XRT light curve. At the time of the UVOT flux rise the X-ray light-curve is characterised by the  fluctuation described in \S \ref{lcx}. This fluctuation,  although statistically significant, is not strong enough to be studied in detail. We cannot exclude that the weak excess at $\sim$45ks, followed by the dip
around $\sim$80ks, has the same origin as the rebrightening observed in the UVOT bands. Unfortunately, the XRT error bars do not allow us to verify this a statement.

\subsection{Late steep decay}
\label{steep}
The decay phase  observed after the break at 225 ks is characterised by a steep index $\alpha_{\rm steep}\approx 2.3$ that,  when considering the contributions of the host galaxy in the optical band  and that of the separate unrelated source in the XRT light curve, describes the evolution in all bands well except for the NIR, where no proper host subtraction is possible. This result, together with the fact that the broadband SED at 450 ks is well described by a single power-law,  suggests a unique mechanism that produces the observed radiation from the NIR to the X-rays.

\section{Possible interpretations of the late rebrigtening}
\label{bump2}
\subsection{One-component scenarios}
As extensively discussed in Nardini et al. (2011), several interpretations have been proposed in the literature to explain the  late-time optical rebrightenings in long GRB afterglows. Some of these interpretations, for instance, the presence of a jump in the external medium density profile (Lazzati et al. 2002, Dai \& Wu 2003, Nakar \& Piran 2003) or a discrete episode of energy injection into the fireball (J\'ohannesson et al. 2006; Fan \& Piran 2006; Covino et al. 2008; Rossi et al. 2011),  assume that the whole broad-band light curve is emitted by the same emitting region without requiring two separate components (see also De Pasquale et al. 2012).
\subsubsection{Density-profile discontinuity}
In these scenarios, the steepness of the second component's rise would be smaller than found by our two-component model in \S \ref{bblc}. In this case we obtain from the fit  $\alpha_{\rm reb}\sim -2$ instead of -2.9, since in the former case there is no contribution at later times from an additional, early component. This value,  makes the rebrightening still too sharp to be explained solely by a jump in the density profile however (Nakar \& Granot 2007; van Eerten et al. 2009). Sharper optical bumps can be obtained  if an evolution of the micro-physical parameters, such as $\varepsilon_b$ and $\varepsilon_e$ in the different density regions, is considered (Kong et al. 2010). However,  a detailed description of how the spectral shape evolves according to density jump models with evolving parameters is lacking in the literature. It is therefore still unclear whether the spectral evolution observed during the rebrightening can be reproduced in this theoretical framework. In particular, unlike the case of GRB 081029 where, there was no evidence of X-ray spectral evolution, in  GRB 100814A we observe both a hint of rebrightening and a simultaneous  $\beta_{\rm X}$ change.  

\subsubsection{Energy injection}
To explain this rebrightening with an energy injection in the fireball,  a sudden release of a large amount of energy at late times is needed.
We have claimed the possible presence of a continuous energy injection in \S \ref{shallow}  to explain the shallow decay-phase after the afterglow onset,  but since the peak flux of the optical bump is almost six times  brighter than the shallow phase light curve extrapolation, the amount of energy injected should be of the same order as the initial afterglow energy release. An energy injection episode is  not expected to produce a change in the observed synchrotron spectral slopes under the standard assumption of non-evolving micro-physical parameters, and thereforedisagrees with the sudden spectral change that occurrs during the bump. This claim is strengthened by the evidence that the break frequency between the optical and X-ray bands was not evolving during the shallow decay-phase and, therefore, the spectral evolution cannot be ascribed to a standard evolution of a synchrotron cooling frequency decreasing in time during the rebrightening.

\subsection{Multi-component scenarios}
\subsubsection{Double-jet}
A different approach we can follow is to consider the two phases of the observed light curve as produced by the sum of two separate components, as empirically represented with the two-component fitting shown in Fig. \ref{fittotale}.  A commonly invoked scenario in cases where a complex chromatic evolution of the afterglow is observed are two separate jets characterised by different opening angles and initial $\Gamma$ Lorentz factors (e.g. Berger et al. 2003, Racusin et al. 2008, Filgas et al. 2011b, de Pasquale et al. 2009).  The early light curve is dominated by the emission of a faster and narrower jet, while a slower and wider second jet becomes dominant at later time because its afterglow onset peaks while the first jet is already decaying fast in its post-jet-break phase. Unlike the extreme cases of GRB 081029 (Nardini et al. 2011) and GRB 100621A (Greiner et al. 2013), the value of $\alpha_{\rm reb}$ is consistent with the pre-onset  rise of a standard afterglow in a homogeneous external medium viewed off-axis  and is also marginally consistent with the on-axis case ($\alpha=-2$) when the uncertainties on the jet-break time of the first component are taken into account.  If this second jet is considered independently from the first one and is assumed to be expanding in an unshocked ISM medium, we can estimate the Lorentz factor of this second jet  to be  $\Gamma_0=16$ following Eq. 15 from Ghirlanda et al. (2011).  This is the preferred model chosen by Liang et al. (2013) to explain the afterglow rebrightenings in their sample.

During the late steep-decay a unique power-law fits all observed bands and we can assume that the $\nu_{\rm c}$ is located at energies below the observed $K_{\rm s}$ band. Applying the standard closure relations (see e.g. Racusin et al. 2009), we see that the observed values of  $\alpha_{\rm steep}=-2.3$ and $\beta=1.08$ found in the fourth SED at 450 ks  agree with the predictions for a uniform spreading jet in a slow cooling regime where for $\nu>\nu_{\rm c}$ and $p>2$, $\alpha=p=2.3$ and $\beta=p/2=1.15$. Before the break, during the flat phase after the rebrightening,  we have that $\Delta\beta=0.5$. Under this scenario this value implies that at 90 ks $\nu_{\rm opt}<\nu_{\rm c}<\nu_{\rm X}$.  Even if we assume that both a passage of the cooling frequency in the GROND bands and a jet break occur around 250 ks, the pre-break decay index is expected to be much steeper than observed ($\alpha_{\rm flat}=3\beta/2=0.71$ instead of 0.22). The difference between $\alpha_{\rm flat}$ and $\alpha_{\rm steep}$ is therefore in contrast with the standard expectations of the closure relations in the same way as discussed in \S \ref{shallow}.

\subsubsection{Ultrarelativistic shell collision}
Vlasis et al. (2011) considered a particular case of energy injection. Late-time rebrightenings were explained as been caused by to late collisions between two ultrarelativistic shells with different Lorentz factors. The central engine does not switch off after emitting the shell responsible for the observed early-afterglow emission but remains active, ejecting another delayed shell that moves with a  steady velocity in the empty medium swept up by the first ejecta's passage. This delayed ejected matter (whose initial velocity could be even lower than that of the first shell) will reach the termination shock of the first shell that was decelerated by the interaction with the external medium. The collision of these shells re-heats the fireball material, resulting in an observed flux increase. The smaller the jet opening angle, the stronger the flux increase and the steeper the rebrightening rise, as shown in Figs. 4 and 5 of Vlasis et al.. Their 1D numerical simulation is  still not able to predict the broad-band SED however, since it was only intended
to reproduce monochromatic optical and radio light curves under different assumptions and  broad-band SED modelling was beyond their scope. On the other hand, GRB 100814A can be used as an interesting test for this type of model. Because this event is characterised by a long-lasting prompt emission with a strong late-time peak it might provide  a possible starting time of these late shells. For example, the bright flare peaking at $\sim$ 150 s could be the signature of another accretion episode. Moreover, if the early optical afterglow is interpreted as the onset of the afterglow related to the first ejecta, we can infer the value of the initial $\Gamma_0$ Lorentz factor of this early shell. Using  again Eq. 15 from Ghirlanda et al. (2011) with an observed peak time of 640 s, we obtain $\Gamma_0=88$. Following this scenario, the double peak observed in the UVOT and XRT bands between 40 and 150 ks, together with the GROND spectral evolution between 100 ks and 300 ks (Fig. \ref{beta}), may suggest a multiple shell ejection. However, the lack of GROND data around the two sub-peaks does not allow a detailed spectral evolution analysis of these separated episodes.

The theoretical possibility that the central engine might produce a multiple-shell acceleration episode has been claimed several times in the literature, either invoking a delayed accretion of the fall-back material onto the black hole (King et al. 2005, Ghisellini et al. 2007, 2009), or taking into account the possible fragmentation of the accretion disk around the black hole at large radii (e.g., Perna, Armitage \& Zhang 2006). From the observational point of view, long periods of quiescence of the central engine, lasting up to thousands of seconds, have been observed in a fraction of GRBs showing the so called pre- or post-cursors in their prompt gamma-ray emission phase (Ramirez-Ruiz et al. 2001; Romano et al. 2006; Burlon et al. 2008, 2009; Gruber et al. 2010). If these gamma-ray emission episodes are associated with the acceleration of different shells with respect to the main prompt event, these are supposed to interact with the latter ones, probably leading to an observable optical flux rebrightening if the difference in $\Gamma$ is large enough. On the other hand, extremely bright optical bumps are associated with GRBs without signatures of visible gamma-ray post-cursors or bright X-ray flares such as in GRB 100621A (Greiner et al. 2013), leaving the issue unsolved.

\subsubsection{Late prompt model}
An alternative two-component scenario invoking  a long-lasting central engine activity is the late prompt model proposed by Ghisellini et al. (2007).  According to this model, the observed optical, NIR, and X-ray light curves
of GRBs are produced by the sum of two separate components: the standard forward-shock afterglow emission, and a radiation
related to a late-time activity of the central engine sustained by the accretion of the fall-back material that failed to reach
the escape velocity of the progenitor star. Unlike the shell-collision model discussed above, the two separate components are emitted in different regions (i.e. the late prompt component in the proximity of the progenitor where also the gamma-ray prompt photons are emitted and the second one at the forward shock front)  and do not share the same micro-physical context. 
\\
The first and second components of the complete GRB 100814A light curve are characterised by a shallow phase followed by a steeper decline, which  agrees with what 
predict the model.
The chromaticity of the rebrightening with respect to the pre-bump phase also  agrees with the predictions of the late-prompt scenario since in this model a difference in the spectral slope in the two phases arises because the components are produced by completely separate radiative mechanisms. This  also occurs in GRB 081029 described in Nardini et al. 2011 (see the related  paper for a more detailed discussion). However, some inconsistencies with the late prompt model predictions appear while considering the fact that in both phases GRB 100814A 
shows clear peaks after fast rises with $\alpha\leq-2$. This value is much steeper than those obtained for  the late prompt component $\alpha_{\rm fl}$ measured for the Ghisellini et al. (2009) sample and would require an extreme geometrical effect, which makes this scenario unlikely for GRB 100814A.  
The fast rise of the second component can be  accounted for by assuming a
different initial time for the late prompt. This would correspond to  considering a
late reactivation of the central engine, instead of a continuous shell
emission, similarly to what is required for the two-shell collision
scenario. However,   $\alpha_{\rm steep}=-2.3$ is in contrast with the idea that this decay is related to the fall-back scenario, unlike the cases described in Ghisellini et al. (2009) and Nardini et al. (2010).

\section{Conclusions}
GRB 100814A is an interesting example of the fast-increasing family of long GRBs that show prominent late optical rebrightenings. Thanks to one of the richest NIR-to-X-ray light-curve coverage ever available, we had the possibility to use this GRB as a laboratory to test the different models explaining these types of rebrightenings. Even if the steepness of the optical bump is much less extreme than what has been observed in other cases such as GRB 081029 or GRB 100621A, the models assuming a unique ejecta emission from the central engine fail in describing the broad-band temporal and spectral evolution we observed. The standard fireball model that describes the broad-band afterglow emission with synchrotron emission from the external shocks  also fails to account  for  the prominent optical rebrightening and  the significant spectral evolution observed in GRB 100814A.  Even  more complex versions of the synchrotron-external shock models with an evolution of the microphysical parameters are not detailed enough to be considered a convincing solution of the problem.   In \S \ref{sed}, \ref{lcphases}, \ref{bump} we discussed in detail the inconsistencies between the predictions of the standard model with the light curve and SED evolution observed during the pre- and post-rebrightening phases.
This weakness of the standard GRB model in describing the growing number of well-sampled GRB afterglow is indeed a well-known topic and has been widely discussed in recent literature, where several alternative models have been proposed. Here we discussed the nature of GRB 100814A in the framework of the most often invoked alternative scenarios, also  comparing this event with other examples of GRBs showing prominent optical rebrightenings without strong  simultaneous features in the X-rays such, as GRB 081029 and GRB 100621A. Similarly  to what was reported in Nardini et al. (2011) and Greiner et
al. (2013), GRB 100814A is also better described by models that invoke more than one component to explain the observed broad-band
light curve evolution.  A more convincing description  can be obtained by considering the possible effect of late-time reactivation of the central engine. We tested different theoretical models, invoking a long-lasting central engine activity, and  found that the most convincing scenario explains the afterglow rebrightening as caused by to the energy injection into the fireball associated to one or more late-time collisions between relativistic shells emitted by the central engine at different times and with different Lorentz factors (as described in Vlasis et al. 2011).  However, the theoretical modelling of radiation processes in this scenario is still not advanced enough to compare it with the detailed spectral data now available. As such, these explanations are  not conclusive. In particular, the most challenging common feature that characterises all GRBs showing prominent optical rebrightenings (such as GRB 100621A, GRB 081029) is the absence of a simultaneous bump in the X-rays. All these events show some hint of X-ray variability around the optical peak time, but the amplitude of the variation is always much smaller than what is observed at longer wavelengths. On the other hand, the late-time
broad-band evolution is consistent with the scenario of a single component connecting the
whole spectrum from the NIR to the X-ray bands. This implies that the
description of the pre-bump X-ray flux of these peculiar events
requires an even more complex scenario.   Neither the shell collision nor the late prompt models, do not consider a similar inconsistency between the optical and X-ray evolution before the rebrightening in their more simplistic versions.  For this reason, this family of events, showing late optical bumps, can be considered as an interesting laboratory for  developing a more detailed modelling of the radiative mechanisms in these scenarios.

\begin{acknowledgements}
We would like to thank the anonymous referee for the useful comments.
ANG and DAK  are grateful for travel funding support through MPE.
FOE acknowledges funding of his Ph.D. thesis through the Deutscher Akademischer
Austauschdienst (DAAD).  Part of the funding for GROND (both hardware and personnel)
was generously granted from the Leibniz-Prize to G. Hasinger (DFG
grant HA 1850/28-1). A. Rossi acknowledges support by the Th\"uringer Landessternwarte Tautenburg. SK and ANG acknowledge support by DFG grant Kl 766/16-1, SS acknowledges support by the Th\"uringer Ministerium f\"ur Bildung, Wissenschaft und Kultur under FKZ 12010-514. This work made use of data supplied by the UK Swift
Science Data Centre at the University of Leicester.
\end{acknowledgements}

\end{document}